\documentclass[a4paper,12pt]{article}
\begin{document}
\hspace{11cm} GEF-TH-01/2001
\vspace {1.5cm}
\begin{center}
{\bf A relation between the charge radii of \boldmath$\pi^{+}$, 
\boldmath$K^{+}$, 
\boldmath$K^{0}$\\ derived by the general QCD parametrization} 
\vskip 30 pt
G.Dillon and G.Morpurgo
\end{center}
Universit\`a di Genova
and Istituto Nazionale di Fisica Nucleare, Genova, Italy.

\vskip 40 pt
\noindent {\bf Abstract.}
We derive, using the general QCD parametrization, the approximate 
relation  $r^{2}(\pi^{+})-r^{2}(K^{+})\cong -r^{2}(K^{0})$ where the $r$'s 
are the charge radii. The relation is satisfied but the experimental 
errors are still sizeable. The derivation is similar to (but even 
simpler than) that for  $r^{2}(p)-r^{2}(n)\cong r^{2}(\Delta^{+})$ 
(Phys. Lett. B 448, 107 (1999)).
\\(PACS: 12.38.Aw, 13.40.Gp ,14.40.Aq)

\vskip 40pt 
\baselineskip 24pt
\noindent {\bf 1. Introduction}
\vskip 5 pt

It has been shown in a series of papers \cite{m,dm} that one can  
parametrize exactly several hadron properties using only general features of 
QCD. The method, named the ``general parametrization'' (GP), allows to 
write, almost at first sight, the most general expression for the 
spin-flavor structure of quantities relevant to the lowest baryons 
(octet+decuplet) and mesons. It turns out that the coefficients of the 
various terms in the parametrization decrease with increasing complexity 
of the term; this is called the hierarchy of the parameters [1g,2a], and 
is the reason why the naive non relativistic quark model \cite{m2}
(NRQM), that takes into account only the simplest terms, works 
reasonably well, in spite of the fact that the motion of the light quarks 
inside hadrons is relativistic. Indeed the GP method was conceived 
[1a] with the aim of explaining the fair success of the NRQM; it 
emerged that, besides explaining this, the method leads to predictions 
in many cases, due precisely to the hierarchy. The GP, though non 
covariant, is relativistically correct, being an exact 
parametrization of a fully relativistic field theory. It is also 
independent [2a] of the choice of the QCD quark mass renormalization 
point.

In a previous paper [2f], we applied the GP to the electric charge 
radii of $p$, $n$ and $\Delta^{+}$ stimulated by a paper of Buchmann,
Hernandez and Faessler \cite{bu1}  who derived the relation   
$r^{2}(p)-r^{2}(n) = r^{2}(\Delta^{+})$ using a quark model including 
two body gluon and pion exchange. We showed, indeed, that such 
relation is obtained quite generally by the GP method, if one neglects 
three index terms and the cloosed loop contribution (absent in 
\cite{bu1}); the expected order of magnitude of the modification due 
to the three index terms was found ($10\%$ to $20\%$)$\cdot  r^{2}(\Delta^{+})$.

Here we apply the GP method to the charge radii of $\pi^{+}$, 
$K^{+}$, $K^{0}$. The calculation is even simpler than that for $p$, $n$ 
and $\Delta^{+}$ because we are now dealing with mesons. Also, while 
the $p$, $n$, $\Delta^{+}$ relation cannot be checked experimentally, 
so far, because the radius of the $\Delta^{+}$ is not known, in the 
present case values for the meson radii, even if affected by large 
errors, exist \cite{ex1, ex2}. In $(fm)^{2}$ it is:
\begin{equation}
	r^{2}(\pi^{+})=0.44\pm 0.01 \quad ; \quad
	r^{2}(K^{+})=0.34\pm 0.05 \quad ; \quad
	r^{2}(K^{0})=-0.054\pm 0.026
	\label{1}
\end{equation}
 \vskip 40pt
 \noindent {\bf 2. The general parametrization of
  \boldmath$r^{2}(\pi^{+})$ , \boldmath$r^{2}(K^{+})$ , \boldmath$r^{2}(K^{0})$}
\vskip 5 pt

The square radius $r^{2}(M)$ of a meson $M$ with e.m. form factor 
$F(q^{2})$ is:
\begin{equation}
	r^{2}(M)=-6\frac{dF(q^{2})}{dq^{2}}\Big\arrowvert _{ q^{2}=0} \quad ; \quad
    F(q^{2})=\langle M(\mbox{\boldmath $q$}/2) \vert \rho (0)
	\vert M(-\mbox{\boldmath $q$}/2)\rangle
	\label{2}
\end{equation}
Here $\vert M(\mbox{\boldmath $p$})\rangle$ is the exact eigenstate 
of the QCD Hamitonian for 
the meson $M$ with total momentum {\boldmath $p$}, 
$\rho(0)=i\bar{\psi}(0)Q\gamma_{4}\psi(0)$, with $\psi (x)$ the 
quark field, and $Q=(1/2)(\lambda_{3}+\frac{1}{3}\lambda_{8})$ is the 
charge operator. Incidentally recall that the $q^{2}$ dependence of the e.m. form 
factors was discussed for $p$ and $n$ in ref.[2g].

Using the technique described in [1,2] (see [1a] or, more brifley, 
the Appendix of [2a,1i]), the exact $r^{2}(M)$ derived from QCD for a 
meson with an $L=0$ auxiliary state (as the lowest pseudoscalar 
mesons) is written in the GP:
\begin{equation}
	r^{2}(M)=\langle W_{M}\vert ``parametrized\ r^{2}"\vert W_{M}\rangle
	\label{3}
\end{equation}
where $W_{M}$ are the standard spin-flavor functions of the $\pi, K$ 
mesons constructed in terms of the quark-antiquark spin flavor 
variables; $W_{M}$ is the product of a singlet quark-antiquark spin 
factor times an octet flavor factor. The notation for mesons is 
similar to that in refs.[1c,1d], except that the projector 
$\Pi^\lambda$ in [1d] is now called $P^s$ ($P^s$ is $1$ acting on a 
strange $q$ or $\bar{q}$ ; $0$ otherwise).

In view of the linearity of $r^{2}$ in $\rho (0)$ the most 
general ``$parametrized\  r^{2}$" for $\pi^{+}$, $K^{+}$, and $K^{0}$ 
is a scalar linear in the quark charges $Q_{i}$ ($Q_{i}=Q_{1},Q_{2}$; note: 
$1=quark$, $2=antiquark$). As to the the spins, the ``$parametrized\  
r^{2}$" can only contain ($\mbox{\boldmath$\sigma$}_{1}\cdot 
\mbox{\boldmath$\sigma$}_{2}$), which, applied to the spin singlet 
factor in $W_{M}$, is just $-3$. Thus the most general
 ``$parametrized\  r^{2}$" (a scalar under rotations, of 
 course) is:
 \begin{equation}
  	 ``parametrized\  
  	 r^{2}"=A\sum_{i}Q_{i}+B\sum_{i}Q_{i}P^s_{i}+C\sum_{i\neq 
  	 k}Q_{i}P^s_{k}+DTr[QP^s]
  	\label{4}
  \end{equation}
  In (\ref{4}) $A,B,C,D$ are four real parameters; the sums are on 
  $i,k=1,2$; for instance, $\sum_{i\neq 
  k}Q_{i}P^s_{k}=Q_{1}P^{s}_{2}+Q_{2}P^s_{1}$ ;\quad for $K^{+}$, it is 
  $\langle W_{K^{+}}\vert Q_{1}P^{s}_{2}+Q_{2}P^s_{1}\vert 
  W_{K^{+}}\rangle = +(2/3)$.
  
  Neglecting the $Trace$ term in (\ref{4}) because, as usual [2a], it 
  is associated to closed internal loops, depressed by the Furry 
  theorem at least 50 times with respect to the dominant term $A$, the 
  eq.(\ref{4}) contains three parameters $A,B,C$ and must fit three 
  quantities. The hierarchy leads to $\vert B/A\vert$=the flavor 
  reduction factor $\cong 1/3$ (more precisely $0.3$ to $0.33$). 
  For $\vert C/B\vert$ (the reduction factor due to one more gluon 
  exchange) we also take $\vert C/B\vert\approx 1/3$ (but here the 
  $1/3$ is less ``universal"; that is, it depends in part on the quantity 
  under consideration and may vary, say, between $0.2$ and $0.37$). 
  In conclusion we set:
  \begin{equation}
    	C\approx (1/3)B\cong (1/9)A
    	\label{5}
    \end{equation}
	 \vskip 40pt
 \noindent {\bf 3. Discussion of the parametrized expression (4) for 
 the radii}
\vskip 5 pt

The eq.(4) gives (neglecting, as stated, the $Trace$ term):
\begin{equation}
	\begin{array}{l}
	r^{2}(\pi^{+})=A\quad (=0.44\pm 0.01) \\
	r^{2}(K^{+})=A+(1/3)B+(2/3)C \quad (=0.34\pm 0.05) \\
	r^{2}(K^{0})=(1/3)B-(1/3)C \quad (=-0.054\pm0.026)
	\end{array}
	\label{6}
\end{equation}
(Incidentally, omitting $C$, the eqs.(\ref{6}) coincide with those of 
the NRQM in its most naive form. Note that, omitting $C$, we obtain 
from (\ref{6}) $r^{2}(K^{0})=-0.010\pm 0.05$.) In general the eqs. 
(\ref{6}) lead to the relation mentioned in the abstract:

  \begin{equation}
    \begin{array}{lcl}
    	    		r^{2}(\pi^{+})-r^{2}(K^{+}) & = & -r^{2}(K^{0})+C  \\
    	    		0.10\pm 0.05 &  & 0.054\mp 0.026\pm 0.05
    	    	\end{array}
    	\label{7}
    \end{equation}
In the numbers below the relation (7), the last $0.05$ is $C$, calculated 
using the hierarchy (\ref{5}). One can see that the GP prediction 
(\ref{7}), also here, agrees with the facts, although more precise 
measurements of $r^{2}(K^{+})$ and $r^{2}(K^{0})$ are needed. 
Incidentally, it is, from (\ref{6}):
\begin{equation}
	\vert B/A\vert =\vert 
	r^{2}(K^{+})-r^{2}(\pi^{+})+2r^{2}(K^{0})\vert /r^{2}(\pi^{+})=\vert 
	0.10\pm 0.05+0.108\mp 0.052\vert /0.44
	\label{8}
\end{equation}
As stated, the hierarchy leads us to expect unambigously $\vert 
B/A\vert \cong 1/3$. We note that, to obtain this value from 
eq.(\ref{8}), it is necessary that the two errors in (\ref{8}) combine 
to $-0.06$; the central values alone produce a too large $\vert 
B/A\vert$.

In conclusion two new results emerge: a) The predicted relation (7) between 
the $r^{2}$'s of  $\pi^{+}$, 
$K^{+}$, $K^{0}$ satisfied as indicated; b) The prediction 
(eq.(8)) that either $r^{2}(K^{+})$ or $-\vert r^{2}(K^{0}) \vert $ (or both)
 must have their center values somewhat larger than the present
ones. 

%
\pagebreak

\end{document}